\documentclass{aa}
\usepackage{booktabs}
\usepackage{array}
\usepackage{supertabular}
\usepackage{graphicx,multicol}
\usepackage{subfigure}
\usepackage{multirow}
\usepackage{setspace}
\usepackage{txfonts}
\newcommand{\placefigAlphaSummary}{
\begin{figure}[htb]
    \centering
    \includegraphics[width=\hsize]{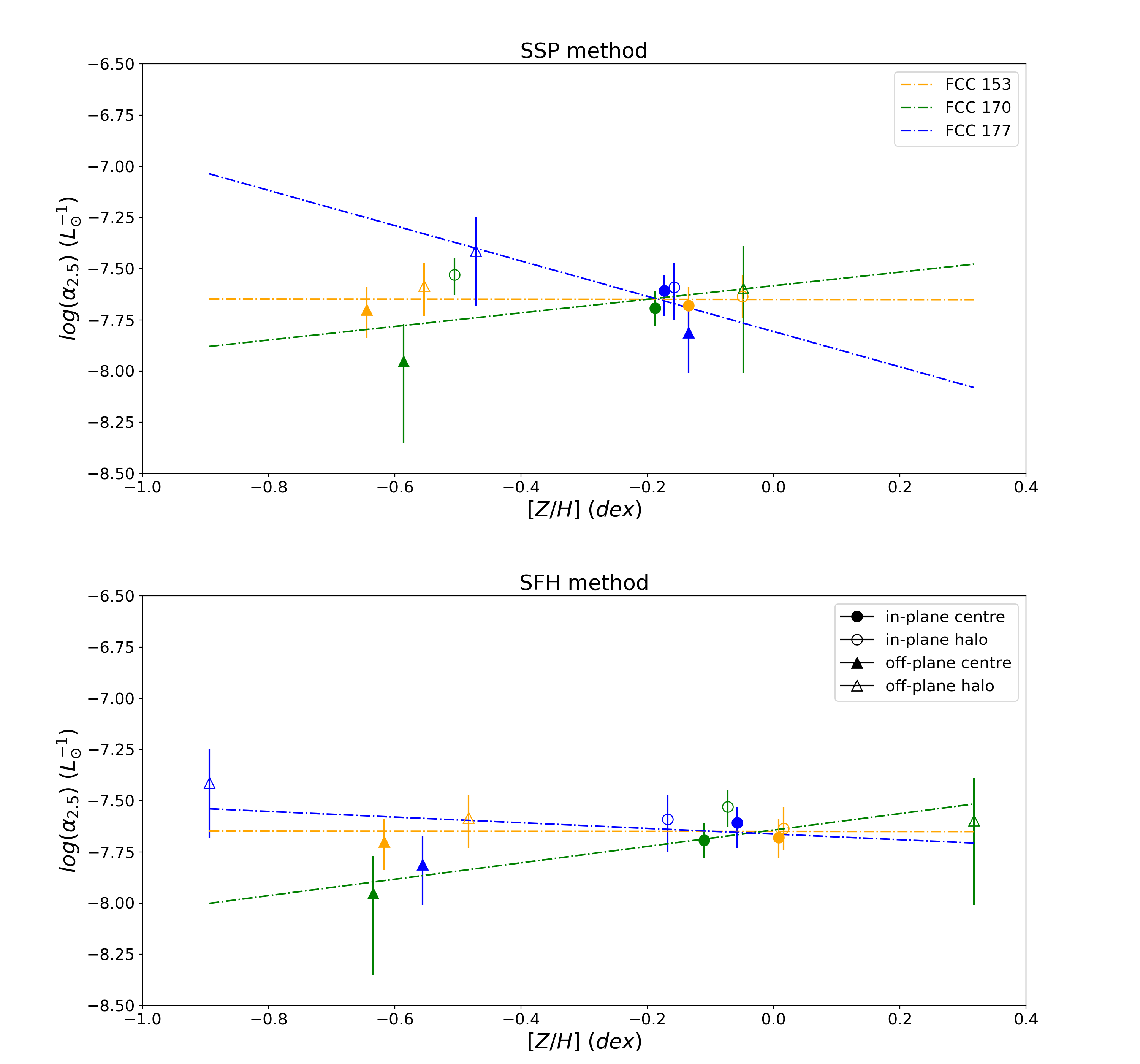}
    \caption{Luminosity specific PNe number $\alpha$ as a function of metallicity for the best-fitting single-age stellar population to selected spectral regions (top panel) following \citet{martin-navarro2019} and for the light-weighted metallicity resulting from a full spectral fitting (bottom panel) following P19.
    Circles and triangles show values from in-plane and off-plane regions, respectively whereas filled and open symbols denote whether these were found within our central or halo MUSE pointings.
    The colour of the symbols indicate measurements from each galaxy, with dashed lines showing linear fits to all values for each object.}
    \label{fig:alpha_summary}
\end{figure}
}

\newcommand{\placefigPNLF}{
\begin{figure}[htb]
    \centering
    \includegraphics[width=\hsize]{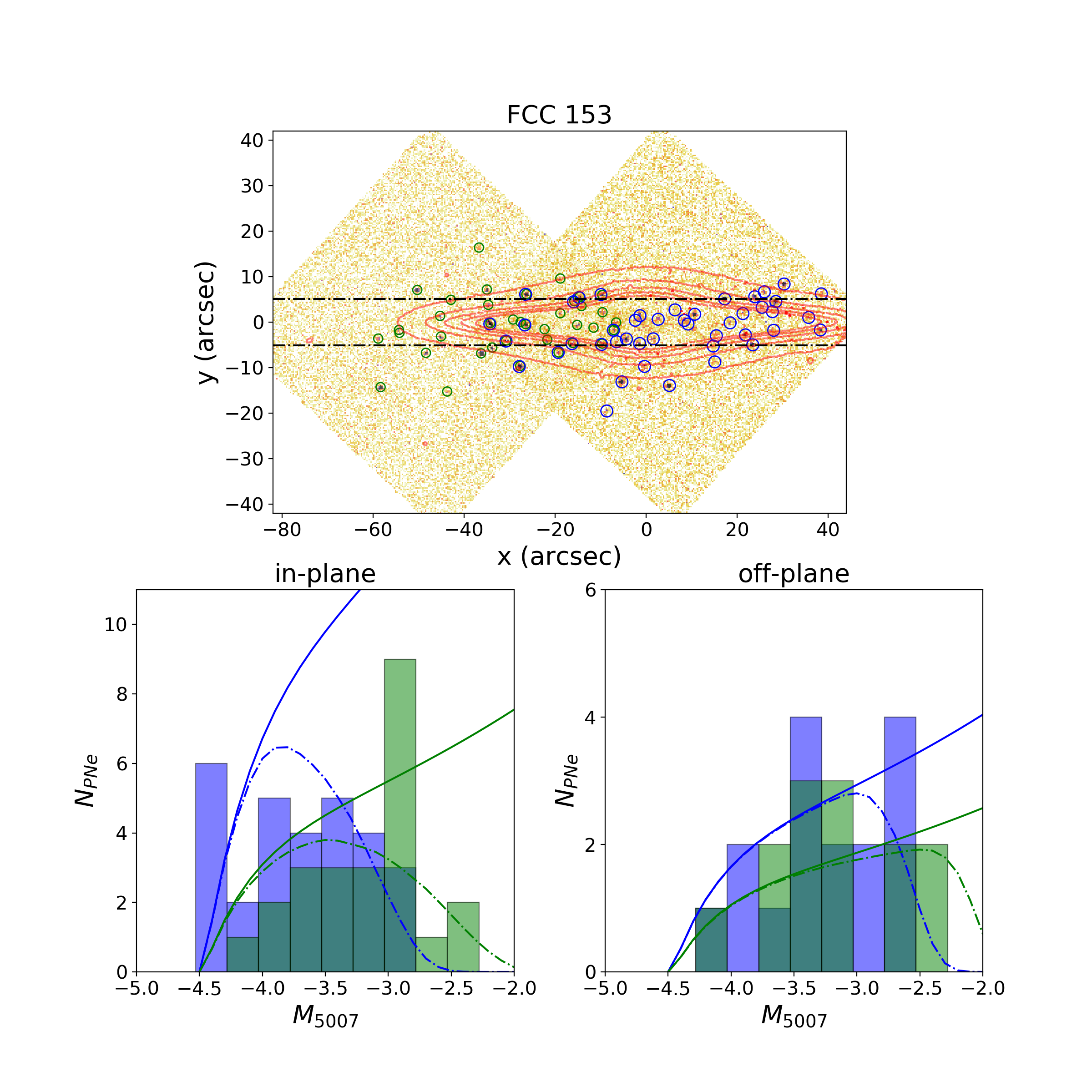}
    \caption{PNLF of FCC\,153 for in-plane and off-plane regions. \textit{Top panel:} map of the [\ion{O}{iii}] $\lambda$5007 line. The blue circles represent the detected PNe in central pointing whereas the green ones show the detected objects in the halo pointing. The black dashed lines indicate the separation between in-plane and off-plane regions. \textit{Bottom panels:} PNLF of both components: in-plane (left panel) and off-plane (right panel). The blue histograms as well as the blue solid and dashed lines represent the PNLF for PNe in the central pointing, whereas the green histograms along with the green solid and dashed lines indicate the PNLF for the PNe in the halo pointing. The red contours are the isophotes of the flux white image.}
    \label{fig:PNLFa}
\end{figure}

\begin{figure}[htb]
    \centering
    \includegraphics[width=\hsize]{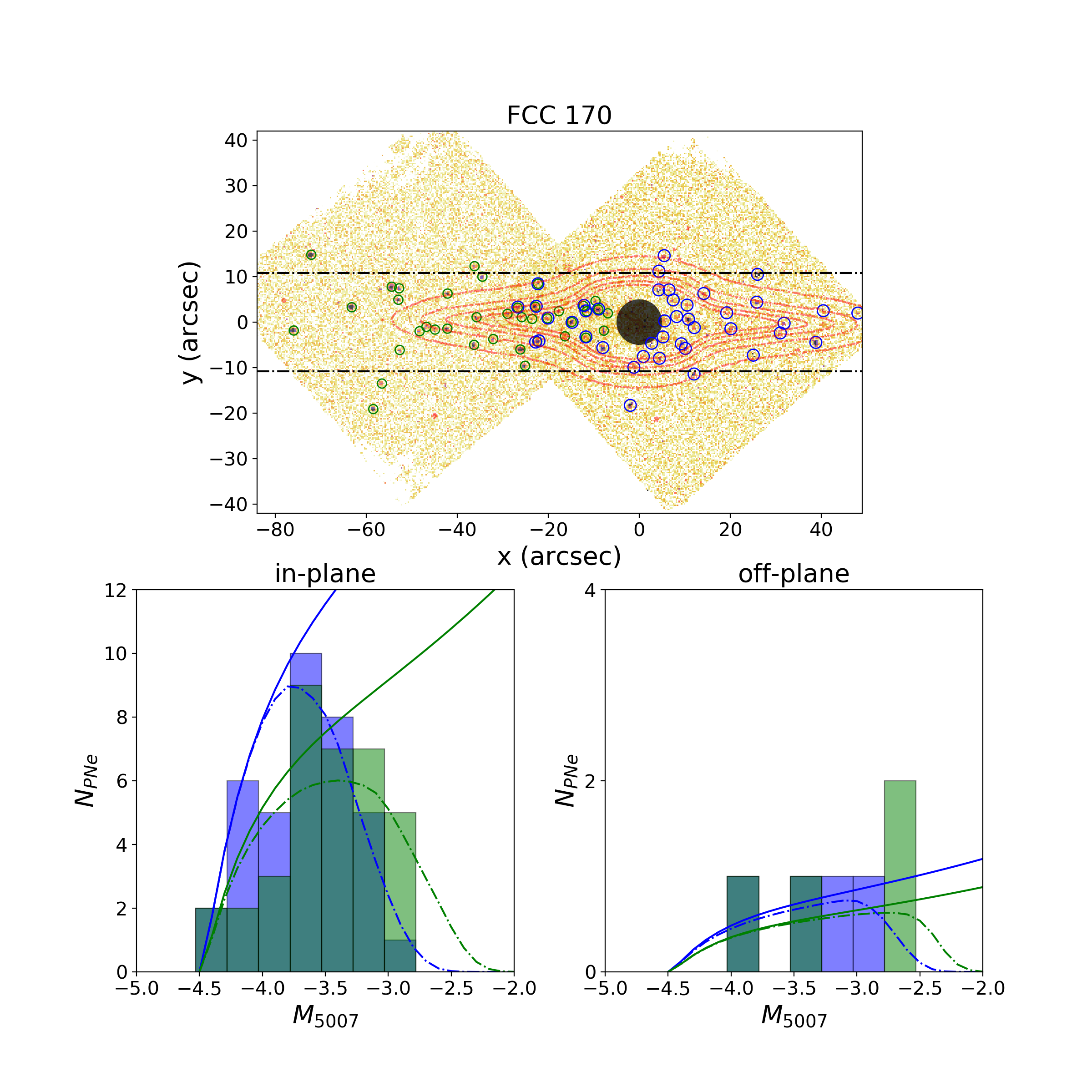}
    \caption{Same as Fig.~\ref{fig:PNLFa}, but for FCC\,170. The black circular area corresponds to the excluded template mismatch region.}
    \label{fig:PNLFb}
\end{figure}

\begin{figure}[htb]
    \centering
    \includegraphics[width=\hsize]{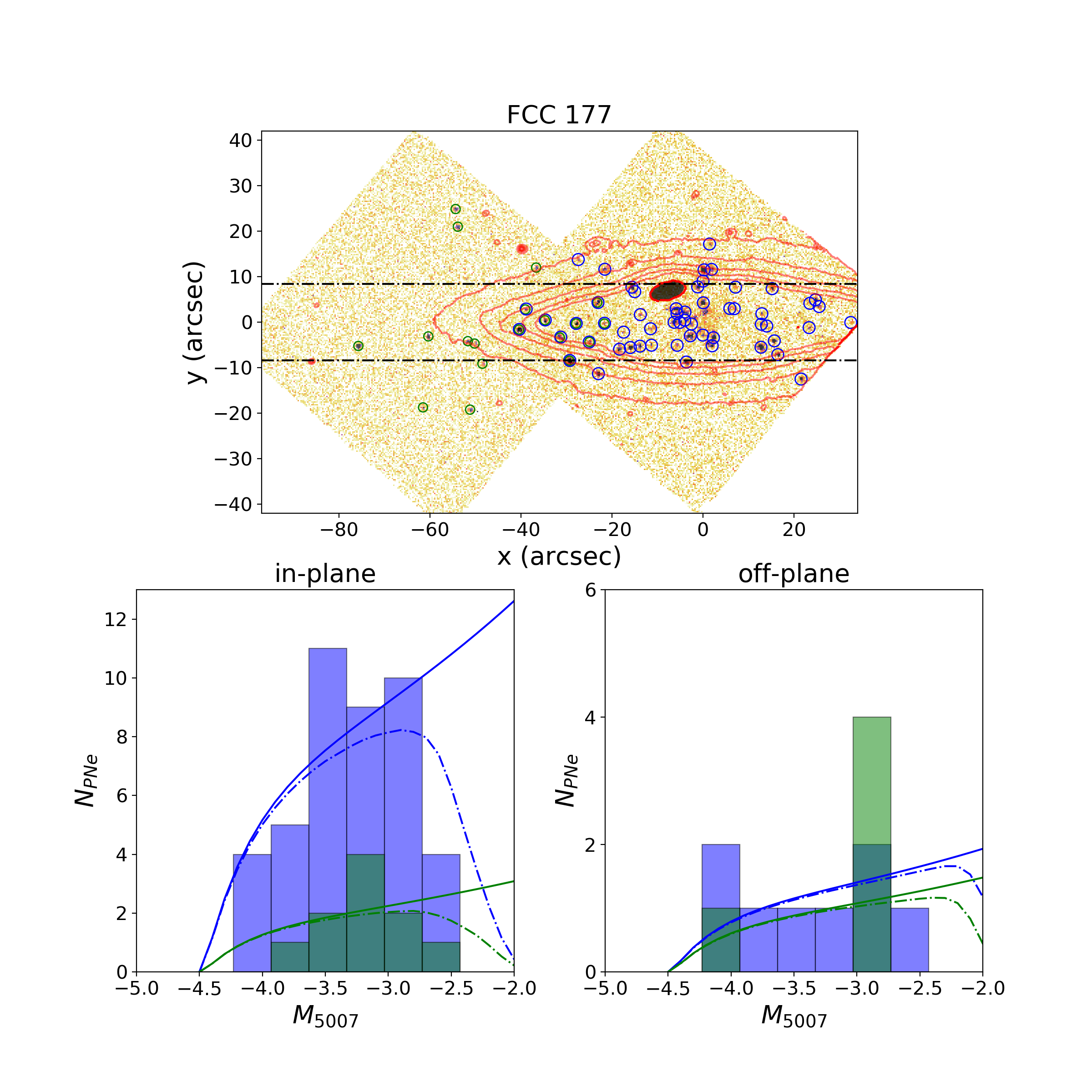}
    \caption{Same as Figs.~\ref{fig:PNLFa}, but for FCC\,177. The black elliptical area corresponds to the excluded foreground galaxy.}
    \label{fig:PNLFc}
\end{figure}
}

\newcommand{\placefigBuzzoni}{
\begin{figure}[htb]
    \centering
    \includegraphics[width=\hsize]{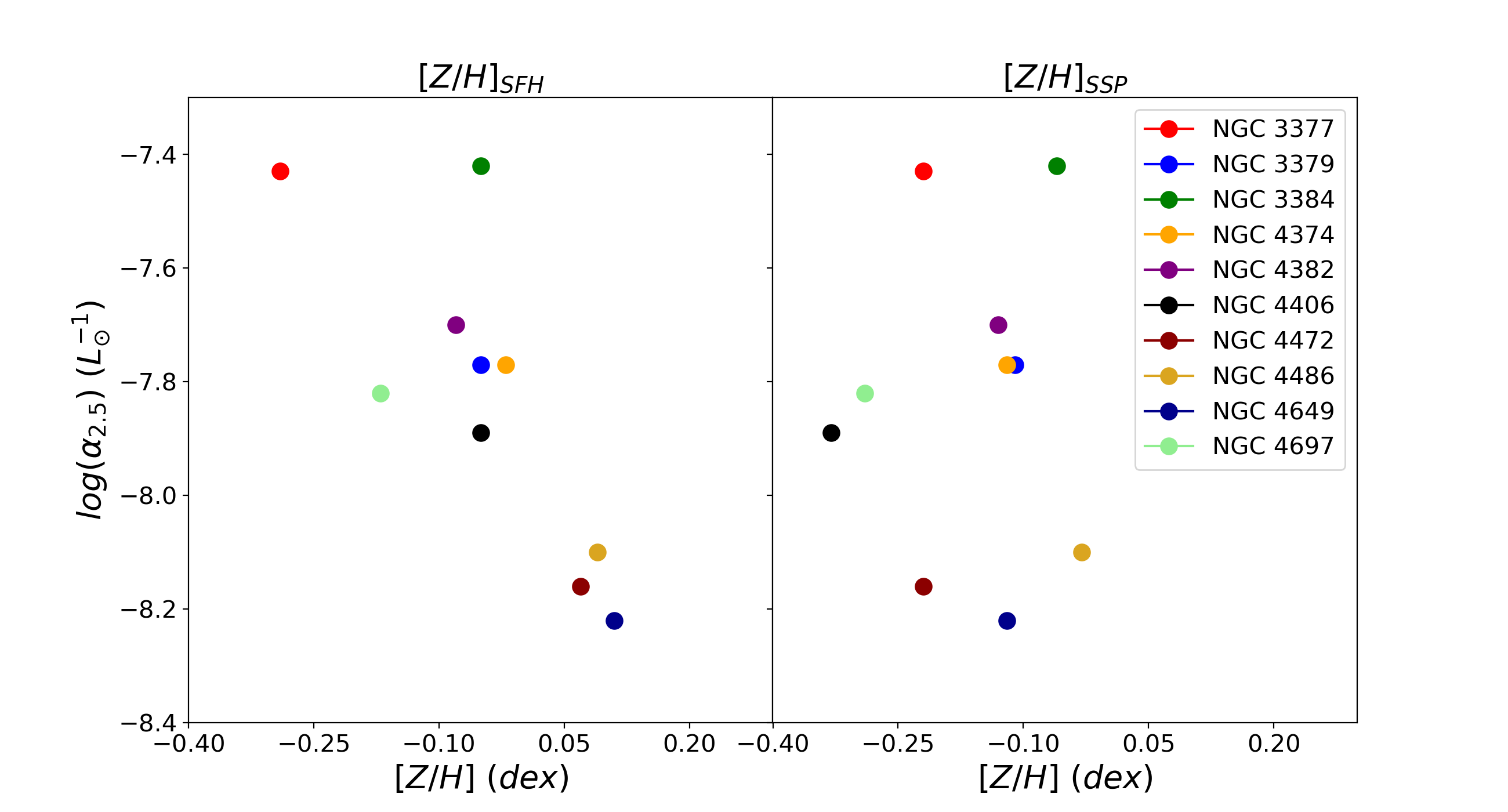}
    \caption{PNe specific number vs stellar metallicity for \citet{buzz2006} sample galaxies covered also by the Atlas$^{3D}$ integral-field spectroscopic survey \citep[see][]{mcdermid2015}, for two different star formation history measurements: full spectral fitting from a single-age stellar population (\textit{left panel}) and full-index-fitting based on a focused set of absorption lines (\textit{right panel}).}
    \label{fig:alpha_buzzoni}
\end{figure}
}
\newcommand{\placetabone}{

\begin{table}
\caption{Point-spread function across the MUSE pointings of the sample galaxies, as characterised using a Moffat function fit to either a foreground star or a selection of bright PNe sources.}
\centering
\begin{spacing}{1.5}
\begin{tabular}{c c c c c}
\hline\hline
Galaxy & Method & Pointing & FWHM & $\beta$ \\
(1) & (2) & (3) & (4) & (5) \\
\hline

   \multirow{2}{*}{FCC 153} & \multirow{2}{*}{PNe} & Centre & $3.6$ & $1.9$ \\
   
    & & Halo &   $3.6$ & $4.2$\vspace{2mm}\\
    
   \multirow{2}{*}{FCC 170} & \multirow{2}{*}{Star} & Centre &  $4.1$ & $2.4$ \\
   
    & & Halo &  $3.5$ & $1.7$ \vspace{2mm}\\
    
   \multirow{2}{*}{FCC 177} & \multirow{2}{*}{PNe} & Centre & $3.8$ & $1.8$ \\
   
    & & Halo & $4.1$ & $4.4$ \\
    
    \hline
\end{tabular}
\end{spacing}
\footnotesize
\textbf{Notes.} (1) galaxy name. (2) selected object to estimate the PSF. (3) selected MUSE pointing. (4) FWHM values are in pixels. (5) $\beta$ parameter of the Moffat profile. The values $FWHM$ and $\beta$ are taken from Spriggs et al. (in preparation).\\
\label{tab:PSF}
\end{table}

}
\newcommand{\placetabtwo}{

\begin{table*}
\caption{Results of the analysis for the MUSE central and halo pointings of the sample galaxies.}
\label{table:central}
\centering
\begin{spacing}{1.3}
\resizebox{\textwidth}{!}{
\begin{tabular}{c c c c c c c c}     

\hline\hline
Galaxy &  Region & PNe detected per region & $N_{2.5}$ & $L_{bol}$ $(L_{\odot})$ & $\alpha_{2.5}$ $(L_{\odot}^{-1})$ & $[Z/H]_{SSP}$ $\rm (dex)$ & $[Z/H]_{SFH}$ $\rm (dex)$ \\
(1) & (2) & (3) & (4) & (5) & (6) & (7) & (8) \\
\hline
   \multicolumn{8}{c}{Central pointing}  \vspace{5mm}\\
   
   \multirow{2}{*}{FCC 153} & in-plane & $29$ & $99^{+22}_{-18}$ & $(4.7^{+0.7}_{-0.5})\times10^{9}$ & $(2.1^{+0.5}_{-0.4})\times10^{-8}$ & $-0.13$ & $0.008$ \vspace{2mm}\\
   
    & off-plane &   $16$  & $24^{+8}_{-6}$ & $(1.2^{+0.2}_{-0.1})\times10^{9}$ & $(2.0^{+0.6}_{-0.5})\times10^{-8}$ & $-0.645$ & $-0.617$ \vspace{5mm}\\
    
   \multirow{2}{*}{FCC 170} & in-plane &  $38$  & $116^{+22}_{-19}$ & $(5.7^{+0.7}_{-0.6})\times10^{9}$ & $(2.0^{+0.4}_{-0.4})\times10^{-8}$ & $-0.188$ & $-0.11$ \vspace{2mm}\\
   
    & off-plane &  $4$   & $7^{+6}_{-3}$ & $(6.3^{+0.8}_{-0.7})\times10^{8}$ & $(1.1^{+0.6}_{-0.7})\times10^{-8}$ & $-0.595$ & $-0.63$\vspace{5mm}\\
    
   \multirow{2}{*}{FCC 177} & in-plane &  $46$   & $62^{+11}_{-9}$ & $(2.6^{+0.5}_{-0.4})\times10^{9}$ & $(2.4^{+0.5}_{-0.6})\times10^{-8}$ & $-0.173$ & $-0.057$ \vspace{2mm}\\
   
    & off-plane &  $10$   & $9^{+4}_{-3}$ & $(5.8^{+1.1}_{-0.9})\times10^{8}$ & $(1.5^{+1.8}_{-1.8})\times10^{-8}$ & $-0.135$ & $-0.556$ \\
    \hline
    
    \multicolumn{8}{c}{Halo pointing} \vspace{5mm}\\

   \multirow{2}{*}{FCC 153} & in-plane & $24$ & $45^{+11}_{-9}$ & $(1.9^{+0.3}_{-0.2})\times10^{9}$ & $(2.3^{+0.6}_{-0.5})\times10^{-8}$ & $-0.049$ & $0.016$ \vspace{2mm}\\
   
    & off-plane &  $14$   & $15^{+5}_{-4}$ & $(5.7^{+0.8}_{-0.6})\times10^{8}$ & $(2.6^{+0.8}_{-0.7})\times10^{-8}$ &  $-0.554$ & $-0.483$ \vspace{5mm}\\
    
   \multirow{2}{*}{FCC 170} & in-plane &   $35$   & $76^{+15}_{-13}$ & $(2.6^{+0.3}_{-0.3})\times10^{9}$ & $(2.9^{+0.6}_{-0.6})\times10^{-8}$ & $-0.51$ & $-0.07$ \vspace{2mm}\\
   
    & off-plane &  $4$   & $5^{+4}_{-3}$ & $(2.0^{+0.3}_{-0.2})\times10^{-8}$ & $(2.5^{+1.5}_{-1.6})\times10^{-8}$ & $-0.048$ & $0.32$ \vspace{5mm}\\
    
   \multirow{2}{*}{FCC 177} & in-plane &  $12$   & $15^{+6}_{-4}$ & $(5.8^{+1.1}_{-0.9})\times10^{8}$ & $(2.6^{+0.8}_{-0.8})\times10^{-8}$ & $-0.158$ & $-0.168$ \vspace{2mm}\\
   
    & off-plane &  $7$   & $7^{+4}_{-3}$ & $(1.8^{+0.3}_{-0.3})\times10^{8}$ & $(3.9^{+1.8}_{-1.8})\times10^{-8}$ & $-0.47$ & $-0.894$ \\
    
    \hline
\end{tabular}
}
\end{spacing}
\footnotesize{\textbf{Notes.} (1) target galaxy. (2) studied region. (3) number of observed PNe. (4) theoretical expected number of PNe down to 2.5 magnitudes. (5) bolometric luminosity. (6) luminosity specific PNe number. (7) light-weighted metallicity derived from SSP index measurements. (8) light-weighted metallicity obtained from the analysis of full spectral fitting.}\\
\end{table*}
}

\begin{document}

   \title{The Fornax 3D project: PNe populations and stellar metallicity in edge-on galaxies}

   \subtitle{}

   \author{P. M. Gal\'an-de Anta
          \inst{1,2}\thanks{pgalandeanta01@qub.ac.uk}
\and
          M. Sarzi\inst{1,3}
\and
          T. W. Spriggs\inst{3}\thanks{tspriggs@outlook.com}
\and
          B. Nedelchev\inst{1,3}
\and
          F. Pinna\inst{4}
\and
          I. Martín-Navarro\inst{4,5,6}
\and
          L. Coccato\inst{7}
\and
          E. M. Corsini\inst{8,9}
\and
          P. T. de Zeeuw\inst{10,11}
\and
          J. Falcón-Barroso\inst{5,6}
\and
          D. A. Gadotti\inst{7}
\and
          E. Iodice\inst{12,7}
\and
          R. J. J. Grand\inst{13}
\and
          K. Fahrion\inst{7}
\and
          M. Lyubenova\inst{7}
\and
          R. M. McDermid\inst{14,15}
\and
          L. Morelli\inst{16}
\and
          G. van de Ven\inst{17}
\and
          S. Viaene\inst{18}
\and
          L. Zhu\inst{19}
          }
          
\institute{Armagh Observatory and Planetarium, College Hill, Armagh, BT61 9DG, UK
\and
Astrophysics Research centre, School of Mathematics and Physics, Queen's University Belfast, Belfast BT7 INN, UK
\and
Centre for Astrophysics Research, University of Hertfordshire, College Lane, Hatfield AL10 9AB, UK
\and
Max-Planck-Institut f\"{u}r Astronomie, K\"{o}nigstuhl 17, 69117 Heidelberg, Germany
\and
Instituto de Astrofísica de Canarias, Vía Láctea s/n, 38205 La Laguna, Tenerife, Spain
\and
Depto. Astrofísica, Universidad de La Laguna, Calle Astrofísico Francisco Sánchez s/n, 38206 La Laguna, Tenerife, Spain
\and
European Southern Observatory, Karl-Schwarzschild-Stra{\ss}e 2, 85748 Garching bei M\"{u}nchen, Germany
\and
Dipartimento di Fisica e Astronomia ‘G. Galilei’, Università di Padova, vicolo dell’Osservatorio 3, 35122 Padova, Italy
\and
INAF–Osservatorio Astronomico di Padova, vicolo dell’Osservatorio 5, 35122 Padova, Italy
\and
Sterrewacht Leiden, Leiden University, Postbus 9513, 2300 RA Leiden, The Netherlands
\and
Max-Planck-Institut fuer extraterrestrische Physik, Giessenbachstrasse, 85741 Garching bei Muenchen, Germany
\and
INAF - Astronomical Observatory of Capodimonte, Salita Moiariello 16, 80131, Naples, Italy
\and
Max-Planck-Institut für Astrophysik, Karl-Schwarzschild-Straße 1, 85748 Garching bei München, Germany
\and
Department of Physics and Astronomy, Macquarie University, Sydney, NSW 2109, Australia
\and
ARC Centre of Excellence for All Sky Astrophysics in 3 Dimensions (ASTRO 3D), Australia
\and
Instituto de Astronom\'ia y Ciencias Planetarias, Universidad de Atacama, Avenida Copayapu 485, Copiap\'o, Chile
\and
Department of Astrophysics, University of Vienna, Türkenschanzstrasse 17, 1180 Vienna, Austria
\and
Sterrenkundig Observatorium, Universiteit Gent, Krijgslaan 281, 9000 Gent, Belgium
\and
Shanghai Astronomical Observatory, Chinese Academy of Sciences, 80 Nandan Road, Shanghai 200030, China
}

   \date{Received XXXXXX; accepted XXXXXXX}

\abstract
   {
   Extragalactic Planetary Nebulae (PNe) are useful distance indicators and are often used to trace the dark-matter content in external galaxies. At the same time, PNe can also be used as probes of their host galaxy stellar populations and to help understanding the later stages of stellar evolution. Previous works have indicated that specific number of PNe per stellar luminosity can vary across different galaxies and as a function of stellar-population properties, for instance increasing with decreasing stellar metallicity.
   }
   {
   In this study we further explore the importance of stellar metallicity in driving the properties of the PNe population in early-type galaxies, using three edge-on galaxies in the Fornax cluster offering a clear view into their predominantly metal-rich and metal-poor regions near the equatorial plane or both below and above it, respectively .
   }
   {
   Using VLT-MUSE integral-field observations and dedicated PNe detection procedures, we construct the PNe luminosity function and compute the luminosity-specific number of PNe $\alpha$ in both   in- and off-plane regions   of our edge-on systems.
   }
   {
   Comparing these $\alpha$ values with metallicity measurements also based on the same MUSE data, we find no evidence for an increase in the specific abundance of PNe when transitioning between metal-rich and metal-poor regions.
   }
   {
    Our analysis highlights the importance of ensuring spatial consistency to avoid misleading results when investigating the link between PNe and their parent stellar populations and suggest that in passively-evolving systems variations in the specific number of PNe may pertain to rather extreme metallicity regimes found either in the innermost or outermost regions of galaxies.
    }

   \maketitle
%

\section{Introduction}

Planetary Nebulae can be detected in external galaxies out to great distances thanks to their intense [\ion{O}{iii}] $\lambda5007$ nebular emission, emitting as much as $10^3\sim10^{4} L_{\odot}$ \citep[e.g.][]{odell1963,paczynski1970,rose1970final,Gesicki2018}.
This makes PNe useful tracers for the kinematics of their parent stellar populations in the very outskirts of galaxies allowing to constrain their dark matter content \citep[e.g.][]{romanowsky2003,douglas2007,coccato2009,kafle2018}. 
The PNe populations of external galaxies have also been found to have similar PN Luminosity Functions (PNLF) with a common bright-end cutoff that in turn makes PNe reliable cosmic distance indicators \citep[][]{ciard1989,ciard2012}.

As tracers of their parent stellar populations, extragalactic PNe can also help to understand the late stage of stellar evolution in galactic environments different from our Milky Way. In fact, both the detailed shape of the PNLF and the fact that PNe population in different kind of galaxies seems to share the same cutoff, are still not fully understood. For instance, the actual presence of bright PNe in old, passively evolving systems is a long-standing problem \citep[][]{marigo2004,ciar2005} that only recent theoretical development has started to address 
\citep[e.g.][]{Gesicki2018,valenzuela2019}. Likewise, both the origin of the PNLF bright cut-off \citep[e.g.][]{ciard2012} and its slope at fainter regimes are not fully understood (e.g. in star-forming regions, as found in the LMC by \citealp{warren2010} or in the outskirts of early-type galaxies, as shown by \citet{hartke2020}.

At a more basic level, the abundance of PNe should also vary across different kinds of galaxies, as proposed by \citet[][hereafter B06]{buzz2006} also in relation to variations to their overall far-UV flux (the so-called UV-upturn, \citealp{burstein1988}). 
In fact, B06 also showed the first observational evidence for an anti-correlation between the specific number of PNe per luminosity ($\alpha$) and stellar metallicity, finding at the same time that the presence of a UV-upturn also corresponds to less abundant PNe populations (consistent also with previous theoretical work by \citealp{greggio1990}).
It is important to note, however, that the analysis of B06 suffers from a spatial inconsistency in their comparison of PNe and stellar population measurements. Indeed, whereas their $\alpha$ values come from previous halo PNe studies \citep[][]{ciard1989,jacoby1989,Hui1993}, their literature measurements for the Mgb index values (used to trace metallicity) and the UV-upturn pertain to the innermost regions of their sample galaxies.

Nowadays, deep imaging facilities allow us to trace radial variations in $\alpha$ out to the most metal-poor halo regions of galaxies \citep[e.g.][]{bhattacharya2019,hartke2020}. Furthermore, the advent of integral-field spectroscopy can both facilitate the measurement of the stellar metallicity in faint galaxy outskirts \citep[by collecting large aperture spectra, e.g.][]{weijmans2009} and detection of PNe deeply embedded in the central bright regions of galaxies \citep[thanks to a detailed spectral modelling, e.g.][]{sarzi2011,pastor2013,spriggs2020}, where they are otherwise inaccessible to narrow-band imaging or counter-dispersed slit-less spectroscopy \citep[e.g.][]{gerhard2005,douglas2007,ventimiglia2011}.

In this respect, edge-on galaxies can be considered as ideal laboratories to test ideas on the connection between PNe and their parent stellar populations. Indeed, the particular inclination of these galaxies makes it possible to directly compare PNe populations in metal-rich disc regions and their counterparts in the more metal-poor bulge or halo, thus allowing to start redressing the spatial inconsistencies in the B06 analysis. 

Using MUSE integral-field spectroscopic data, in this paper we explore the PNe populations for the three edge-on galaxies FCC\,153, FCC\,170, and FCC\,177. These targets were observed during the magnitude-limited Fornax3D survey for bright galaxies within the virial radius of the Fornax cluster \citep[][]{sarzi2018}. We place our findings in the context of the stellar population measurements previously reported by \citet[][hereafter P19]{pinna2019FCC170,pinna2019b}. These edge-on galaxies host a range of stellar populations that can be divided into a predominantly metal-rich bulge/thin-disc and a metal-poor off-plane populations. In this respect, PNe can be used to trace the kinematics of these parent stellar populations and check whether the abundance of hosted PNe is dependent on different star formation signatures such as the metallicity.

This paper is organised as follows. In Section~2, we present a brief summary of the sample galaxies as well as the properties of the observations. In Section~3, we explain the methodology and procedures to identify and confirm the detected PNe, as well as to estimate the PNLF, light-weighted metallicity, and luminosity specific PNe number. In Section~4, we present and discuss the results of both the PNLF for the three galaxies and specific PNe number per galaxy as function of the metallicity of different in-plane and off-plane regions. Lastly, in Section~5 we give our conclusions.

\section{Observations and data reduction}

The MUSE data for FCC\,153, FCC\,170 and FCC\,177 (IC\,1963, NGC\,1381, and NGC\,1380A, respectively) were obtained in the wide-field mode, which ensures a spatial sampling of $0.2\arcsec\times0.2\arcsec$  on a $1\arcmin\times1\arcmin$ field-of-view (FoV). The wavelength range of MUSE cubes ranges between 4650 \AA\ and 9300 \AA\ with a spectral sampling of $1.25$ \AA\ pixel$^{-1}$ and average instrumental spectral resolution of ${\rm FWHM_{int}}=2.5$ \AA\ \citep{sarzi2018}. In the particular case of our three edge-on galaxies, the MUSE data comprises of a central pointing and offset pointing further covering the outer disc and halo regions, and to which from now on we refer as the halo pointing. Different integration times of 1h and 1h30min for central and halo pointing, respectively, allow to reach the same limiting surface brightness of $\mu_{B}=25 \rm \,mag\,arcsec^{-2}$. Central pointings were required to be observed under good seeing conditions (FWHM $< 0.8\arcsec$), whereas halo pointings had less stringent image quality constraints (FWHM $< 1.5\arcsec$). This will be accounted in our PNe analysis (see Section~3.2). 

\placetabone

The data reduction was performed using the MUSE pipeline \citep[][]{weilbacher2012} applying the ESOREFLEX environment \citep{freudling2013} as described in \citet{sarzi2018} and \citet{iodice2019}. This includes key steps such as sky-subtraction, telluric correction and both relative and absolute flux calibration. Normally, the single pointings would be aligned through reference stars and further combined to produce final MUSE mosaics, but here this last step was skipped since our PNe analysis will need to be separately applied to our individual finally-reduced central and halo pointings due the aforementioned differences between their imaging quality.

Maps for the stellar kinematics of these objects were presented both in \citet{pinna2019FCC170,pinna2019b} and in \citet{iodice2019}, with the former works additionally showed maps for the stellar age and metallicity.
These studies also noticed the absence of diffuse ionised-gas emission in these galaxies, although the presence of PNe in these and other passively-evolving objects was already being mentioned in \citet{iodice2019}. This is  thanks to a careful spaxel-by-spaxel separation of the stellar continuum and nebular emission using the GandALF code \citep[e.g.][]{sarzi2006} within the novel GIST pipeline of \citet{bittner2019}, as also described in \citet{sarzi2018}.
%

\section{Methodology}

In order to characterise the PNe populations of our edge-on galaxies we follow the method of \citet{spriggs2020}. 
This consists of five separate steps: i) identification of PNe candidate using [\ion{O}{iii}] $\lambda$5007 signal-to-noise maps obtained from our dedicated spaxel-by-spaxel GandALF fits, ii) dedicated 3D-fitting of such PNe candidates for their kinematics and total [\ion{O}{iii}] $\lambda$5007 flux while imposing a fixed spatial profile for their [\ion{O}{iii}] $\lambda$5007 emission according to the (pre-determined) spatial Moffat point-spread function (PSF)\footnote{The PSF evaluation assumes a \citet{moffat1969theoretical} profile and is achieved using foreground stars by simultaneously applying our 3D-fitting procedure to the [\ion{O}{iii}] $\lambda$5007 emission of a few bright PNe. The radial extent and kurtosis of the Moffat profile is defined by its parameters $\alpha$ and $\beta$, which relate to the $\rm FWHM = 2\alpha\sqrt{2^{1/\beta}-1}$.}, iii) isolating and removing  PNe interlopers based on the comparison with the host-galaxy stellar kinematics and unresolved HII-regions or supernovae remnant PNe impostors using line diagnostics, iv) construction of the PNLF and of our PNe detection incompleteness function (Section~3.1) and v) finally, estimation of the total luminosity-specific number of PNe within a given magnitude limit (Section~3.2). This last step requires either adopting or deriving the galaxy distance using the PNLF, and involves re-scaling the incompleteness-corrected model for the PNLF to match the observed number of PNe.

The first three steps of this analysis deliver a catalogue for the PNe contained within the FoV of our MUSE observations. Separate PNe catalogues for each pointing are needed in first place since the central and halo data were obtained under rather different seeing conditions (Table~\ref{tab:PSF}), which impacts on the shape and extent of the incompleteness functions that we then fold in the PNLF modelling. Our detected PNe are in agreement with those in Spriggs et al. (in preparation), after that, the only difference is that here we will separately apply the last two steps of our PNe analysis to the  predominantly metal-rich and metal-poor regions near and off the equatorial plane of our edge-on galaxies, respectively. We obtain separate PNLF and luminosity-specific PNe numbers in order to explore variations in the PNe populations of regions with significantly different parent stellar populations. For this we apply the stellar population differentiation adopted in \citet{pinna2019b,pinna2019FCC170}, which is shown by horizontal lines in the top panels of Figs.~\ref{fig:PNLFa}-\ref{fig:PNLFc}.
Finally, using two complementary approaches we will proceed to estimate the stellar metallicity in these regions (Section~3.3) in order to place our PNe population results in the context of the B06 relation, which we will also re-evaluate using the stellar metallicity measurements from the literature (Section~3.4).

\placefigPNLF

\subsection{Observed PNLF, completeness corrections and best-matched PNLF models}

Following the procedure outlined in the previous section we obtain separate catalogues for PNe populations encompassed by our central and halo MUSE pointings. The location of our detected PNe is shown in the top panels of Figs.~\ref{fig:PNLFa}-\ref{fig:PNLFc} for each galaxy, with blue and green circles for PNe identified in the central and halo pointing, respectively. We then further divide the central and halo PNe in two sub-catalogues for regions close to the equatorial plane and those lying above and below it, and construct corresponding PNe luminosity functions as shown in the lower panels of Figs.~\ref{fig:PNLFa}-\ref{fig:PNLFc}.
To first order, after accounting for the incompleteness of our observations, adjusting for distance and applying an appropriate re-scaling, our observed PNLF should compare well with the standard form of the PNLF function introduced by \citet{ciard1989}. This is given by 
\begin{equation}
    N(M)\propto e^{0.307M_{5007}}\left[1-e^{3(M^{\star}_{5007}-M_{5007})}\right],
    \label{ec:PNLF}
\end{equation}
where $M^{\star}_{5007}=-4.52\;\rm mag$  is the characteristic bright-end cut-off \citep[according to the latest calibration of][]{ciard2012} that makes the PNLF a useful distance indicator.

Following \citet{spriggs2020} we define the PNe detection completeness at a given apparent magnitude $m_{5007}$ as the fraction of galaxy stellar light within the MUSE FoV where a PNe of that particular magnitude can be detected.
For this, we apply our PNe detection criteria on a spaxel-by-spaxel basis, by first computing the peak [\ion{O}{iii}] $\lambda$5007 flux for PNe of apparent magnitude $m_{5007}$ (given our PSF model) and subsequently checking if the corresponding peak $A_{[\ion{O}{iii}]}$ spectral amplitude (given the MUSE spectral resolution and the typical PNe intrinsic $\sigma$ of $\sim 40\,\rm km\,s^{-1}$) exceeds three times the local residual-noise level from our previous spaxel-by-spaxel GandALF spectral fitting.

This procedure is equivalent to randomly populating the MUSE FoV with simulated PNe of magnitude $m_{5007}$ while considering that larger numbers would be expected in brighter regions, and then simply evaluate the completeness as the fraction of PNe that would be detected \citep[see also][]{sarzi2011,pastor2013}.
During this procedure we also exclude regions affected by either background or foreground sources (e.g. in FCC\,177) and where we have evidence of significant systematic effects in our spectral fitting (i.e. template-mismatch during the spaxel-by-spaxel GandALF fit) that would lead to spurious PNe detection or biased [OIII] flux measurements (e.g. in the central regions of FCC\,170).

The outlined procedure is separately applied to the in and off-plane regions in both central and halo pointings to construct the PNLFs shown in Figs.~\ref{fig:PNLFa}-\ref{fig:PNLFc}.

With these completeness-correction functions at hand, we first consider the best distance to our sample objects from Spriggs et al. (in preparation), using all the data from the central pointing, where the seeing conditions are less prominent. In short, this procedure adjusts the distance modulus until the observed cumulative PNLF distribution is optimally matched by the corresponding distance-shifted and completeness-corrected cumulative \citeauthor{ciard2012} PNLF, through a minimisation of the Kolgorov-Smirnov statistics. Once this best distance is obtained, we compute appropriately shifted and completeness-corrected model PNLF for each of the in- and off- plane regions in the central and halo pointing, further rescaling them until their integrated value matches the total number of PNe observed in these regions. Such adjusted model PNLFs are also shown in the lower panels of Figs.~\ref{fig:PNLFa}-\ref{fig:PNLFc}, and are based on best-fitting distances of 21.7, 19.6 and 17.8 Mpc for FCC153, FCC170 and FCC177, respectively (Spriggs et al. in preparation). As detailed in that paper, these distances are in good agreement with other measurements, such as those based on surface brightness fluctuations as in \citet{blackeslee2009}.

\subsection{Total and Luminosity-specific PNe numbers}

Once the \citet{ciard1989} PNLF has been matched to the observed PNLF in the in- and off-plane regions of our edge-on galaxies, we can estimate the total number of PNe from the bright cut-off down to a chosen magnitude limit by simply integrating the shifted and re-scaled model PNLF as it is. Following \citet[][]{buzz2006}, in this way we estimate the true number $N_{2.5}$ of PNe 2.5 magnitude from the PNLF bright cut-off. To this extrapolated total, we associate errors obtained from appropriately rescaling the Poisson uncertainty on the  the actual number of PNe that we observe within a given region, computed following the prescription of \citet{gehrels1986}.

From this number, we can estimate the so-called luminosity specific number of PNe
\begin{equation}
    \alpha_{2.5} = \frac{N_{2.5}}{L_{bol}},
    \label{eq:alpha}
\end{equation}
where $L_{bol}$   is the bolometric luminosity for the region where we have been looking for 
PNe. In turn this is given by
\begin{equation}
    L_{bol}(L_{\odot})=10^{-0.4(M_{\lambda}-M_{\lambda,\odot})}\times10^{-0.4(BC_{\lambda}-BC_{\lambda,\odot})},
    \label{eq:bol_lum}
\end{equation}
where $M_{\lambda}$ and $M_{\lambda,\odot}$ are the absolute magnitude for the region of interest within the galaxy and the Sun within a chosen broad-band filter, while
$BC_{\lambda}$ and $BC_{\lambda,\odot}$ are the corresponding bolometric magnitude corrections from the given broad-band filter. See also \citet[][Eq. 9]{Torres2010} and \citet[][Eq. 14]{hartke2020}.

In the case of our MUSE observations, we have chosen the SDSS $g$-band to compute the observed $m_g$ and absolute $M_g$ magnitude of the in and off-plane regions  in our edge-on galaxies. As for deriving the bolometric correction, we follow the procedure described by \citet[][]{spriggs2020} whereby $BC_{g}$ is derived using the optimal combination of EMILES stellar templates \citep[][]{vazdekis2016} that best matches the MUSE integrated spectrum observed within the region of interest (i.e. our in and off-plane  region in both central and halo pointing).
Adopting a bolometric correction for the Sun in the $g$-band of $BC_{g,\odot}=-1.78\;\rm mag$ and a g-band absolute magnitude of $M_{g,\odot}=5.23\;\rm mag$ \citep{willmer2018}, and using Eq.~(\ref{eq:bol_lum}), we estimate the bolometric luminosity of our in and off plane regions, which are reported in Table~\ref{table:central}. Errors in $L_{bol}$ are dominated by $M_{\lambda}$ errors from distance uncertainties.

\placetabtwo

\subsection{Stellar metallicity measurements from integrated MUSE spectra}

In order to place our PNe findings in relation to the properties of their parent stellar population and in particular to check on any dependence of $\alpha_{2.5}$ with stellar metallicity, we proceed to measure the latter in both the in and off-plane regions in and above the disc of our edge-on galaxies, both in MUSE central and halo pointings. For this, we integrate our MUSE spectra in these regions (as defined by P19; Section~3) over a minimum signal-to-noise of 1, obtaining 4 different spectra for each galaxy, and we derive the stellar metallicity according to the following two approaches:

\begin{itemize}
    \item Full spectral fitting using the pPXF code \citep{Cappellari2004,cappellari2017} in the 4750-5500 \AA\ wavelength range as done in P19, using a set of single-age stellar population spectral models from the MILES library \citep[][]{falcon-barroso2011} to estimate an average value for the stellar metallicity. 

    \item Spectral fitting focused on the wavelength regions around a specific set of absorption-line features particularly sensitive to either age (e.g. $H_{\beta}$), metallicity (e.g. \ion{Fe}{5270}, \ion{Fe}{5335}, etc.) or $\alpha$-element abundance (e.g. \ion{Mg}{5177}) as done in \citet{martin-navarro2019} using also the single-age stellar population models from the MILES library to derive a single, best-fitting value for the metallicity $\rm [Z/H]_{SSP}$.
    
\end{itemize}

These two methods are complementary. The spectral-fitting approach of P19 allows to account for an extended star-formation and metal-enrichment history through a combination of single-age models of varying age and metallicity instead of adopting only the values corresponding to the best matching single-age stellar population model as in \citet{martin-navarro2019}. The latter approach however also accounts for more subtle variations in $\alpha$-element abundances (and in fact, also in the slope at the low-mass-end of the stellar initial-mass function), to which full-spectral fitting can be rather insensitive \footnote{Whereas P19 derived mass-weighted values here we use instead light-weighted averages $\rm [Z/H]_{SFH}$ in order to be more consistent with the approach of \citeauthor{martin-navarro2019} Our conclusions are not affected by this choice.}.

\subsection{Revising the \citeauthor{buzz2006} relation}\label{sec:3.4}

\placefigBuzzoni

The evidence for the anti-correlation between the specific number of PNe $\alpha_{2.5}$ and stellar metallicity of B06 was based on literature measurements for the strength of the Mg$_2$ absorption-line index rather than an actual metallicity measurement. To quantify the $\alpha_{2.5}$ variations that they would have found with actual metallicity measurements, we took all the objects in B06 sample that were also observed over the course of the ATLAS$^{3D}$ survey \citep[][]{Cappellari2011}. Using the measurements of \citet{mcdermid2015} we revisit the $\alpha_{2.5}$-metallicity relation of B06. As in the case of our two adopted approaches to estimate the stellar metallicity, \citet{mcdermid2015} provide average (light-weighted) values based on full-spectral fitting $\rm [Z/H]_{SFH}$ and single-age values from the best stellar-population model $\rm [Z/H]_{SSP}$ that best matched the observed strength of a set of absorption line-strength indices, while  accounting for [{$\rm \alpha/Fe$}] variations.

Fig.~\ref{fig:alpha_buzzoni} shows that such a revised $\alpha_{2.5}$ - metallicity relation remains in place only when considering $\rm [Z/H]_{SFH}$ values, indicating at best a factor $\sim$6 variation in the specific number of PNe per 0.4 dex variation in metallicity. 
No visible trend is present when considering $\rm [Z/H]_{SSP}$ values. This suggests that either [{$\rm \alpha/Fe$}] may also play a part in the original trend with the Mg$_2$ line-strength index (which subsists also when looking at the ATLAS$^{3D}$ \ion{Mg$_b$} values) or caution is needed when considering the widely different spatial scales probed by B06.

\section{Results and discussion}\label{sec:4}

Given our final estimates for the specific number of PNe 2.5 magnitudes from the PNLF bright cut-off and stellar metallicity in the regions near or above and below the equatorial plane  of our sample galaxies (Table~\ref{table:central}), Fig.~\ref{fig:alpha_summary} finally shows how these quantities fare against each other across such different regions and as observed both in our central and halo MUSE pointings.

We observe no significant difference in $\alpha_{2.5}$ between in and off-plane regions, irrespective of the methodology used to estimate the stellar metallicity, finding at most a factor $\sim$2.6 increase in $\alpha_{2.5}$ between the off-plane measurements of FCC\,177 where there is a $0.33$ dex difference. This lack of a trend appears to contradict the results of B06. Indeed, even though the metallicity estimates for the predominantly metal-poor off-plane regions can span a wide range of values, on average they are $\sim$0.4 dex lower than their counter parts closer to the equatorial plane and therefore ought to show up to $\sim$6 times higher values for $\alpha_{2.5}$  according to our most optimistic revision of the B06 analysis (Section~3.4). Despite the large uncertainties in the $\alpha_{2.5}$ values for such fainter off-plane regions, there simply is no room to accommodate a factor 6 (i.e. $\sim$0.8 dex) increase in $\log(\alpha_{2.5})$.

This result is at odds with the recent work of \citet{hartke2020}, who find a clear increase in the specific PNe number in the outskirts of the closer Leo Group early-type galaxy NGC 3379 (M105), based on deep narrow-band on-off imaging extending out to nearly 23 effective radii ($R_{e}$). This study reveals, however, that such an enhancement in $\alpha$ values begins from 8$R_{e}$ onward, deep in the outer stellar halo or even in the intra-group light medium and at metallicity regime most likely below the one found in the metal-poor extra-planar regions of our edge-on galaxies. 
On the other hand, at smaller radii \citet{hartke2020} find no $\alpha$ variation despite the presence of large stellar population gradients (e.g. within 3-4 $R_{e}$, \citealp{weijmans2009}) and this is consistent with our findings.
Changing perspective, the lack of a trend between $\alpha$ and metallicity contrasts also with the decrease in the number of bright post-AGB stars and increase in the far-UV flux towards the nuclear metal-rich regions of M31 found by \citet{rosenfield2012}, and corresponds to tentative evidence for a decrease in the number of bright PNe \citep[][]{pastor2013}. 
Although edge-on systems are not particularly suited to explore the innermost and crowded PNe populations of galaxies, future MUSE studies assisted by adaptive optics have the potential to shed more light in this respect.

\placefigAlphaSummary

\section{Conclusions}

Using MUSE observations, we have explored the PNe populations in three edge-on galaxies in the Fornax cluster, offering an unique benchmark for testing the presence of systematic differences in the PNe content between predominantly metal-rich regions near the equatorial plane and generally metal-poor off-plane populations. We have found no significant evidence of a change in the specific
number of PNe between the metal-poor and metal-rich regions of our
edge-on galaxies, despite us probing a range in stellar metallicity
values (0.5 dex) similar to that of previous investigations which
were however comparing the properties of PNe and their parent
stellar population on rather different spatial scales. Presently,
our results further suggest that variations in the specific number
of PNe in passively-evolving systems may pertain either to extreme
metallicity regimes and to the innermost or outermost regions of
galaxies.

\begin{acknowledgements}

This work was supported by Science and Technology Facilities Council [grant number ST/R504786/1]. We really appreciate the computational facilities provided by University of Hertfordshire and Swinburne University to provide us with the necessary storage and analysis of the MUSE data, necessary to carry out our analysis. F.P., I. M-N, and J. F-B acknowledge support through the RAVET project by the grant PID2019-107427GB-C32 from the Spanish Ministry of Science, Innovation and Universities (MCIU), and through the IAC project TRACES which is partially supported through the state budget and the regional budget of the Consejer\'ia de Econom\'ia, Industria, Comercio y Conocimiento of the Canary Islands Autonomous Community. EMC acknowledges support by Padua University grants DOR1885254/18, DOR1935272/19, and DOR2013080/20 and by MIUR grant PRIN 2017 20173ML3WW$_{0}$01.GvdV acknowledges funding from the European Research Council (ERC) under the European Union's Horizon 2020 research and innovation programme under grant agreement No 724857 (Consolidator Grant ArcheoDyn).

\end{acknowledgements}


\bibliographystyle{aa}
\bibliography{bibliography}

\end{document}